\documentclass[prl,twocolumn,showpacs,superscriptaddress,floatfix,amssymb,reprint]{revtex4}
\usepackage{graphicx}

\newcommand{\ket}[1]{| #1 \rangle}

\newcommand{\SIGMA}{\mbox{\boldmath${\sigma}$}}
\newcommand{\NABLA}{\mbox{\boldmath${\nabla}$}}

\begin{document}

\title{Seeing topological order in time-of-flight measurements}
\author{E. Alba}
\affiliation{Instituto de F\'{\i}sica Fundamental, IFF-CSIC, Calle Serrano 113b, Madrid 28006, Spain}
\author{X. Fernandez-Gonzalvo}
\affiliation{Instituto de F\'{\i}sica Fundamental, IFF-CSIC, Calle Serrano 113b, Madrid 28006, Spain}
\author{J. Mur-Petit}
\affiliation{Instituto de F\'{\i}sica Fundamental, IFF-CSIC, Calle Serrano 113b, Madrid 28006, Spain}
\author{J. K. Pachos}
\affiliation{School of Physics and Astronomy, University of Leeds, Leeds, LS2 9JT, United Kingdom}
\author{J. J. Garcia-Ripoll}
\affiliation{Instituto de F\'{\i}sica Fundamental, IFF-CSIC, Calle Serrano 113b, Madrid 28006, Spain}

\date{\today}


\begin{abstract}
In this work we provide a general methodology to directly measure topological order in cold atom systems. As an application we propose the realisation of a characteristic topological model, introduced by Haldane, using optical lattices loaded with fermionic atoms in two internal states. We demonstrate that time-of-flight measurements directly reveal the topological order of the system in the form of momentum space skyrmions.
\end{abstract}

\maketitle

\begin{figure*}
  \centering
  \includegraphics[width=0.87\linewidth]{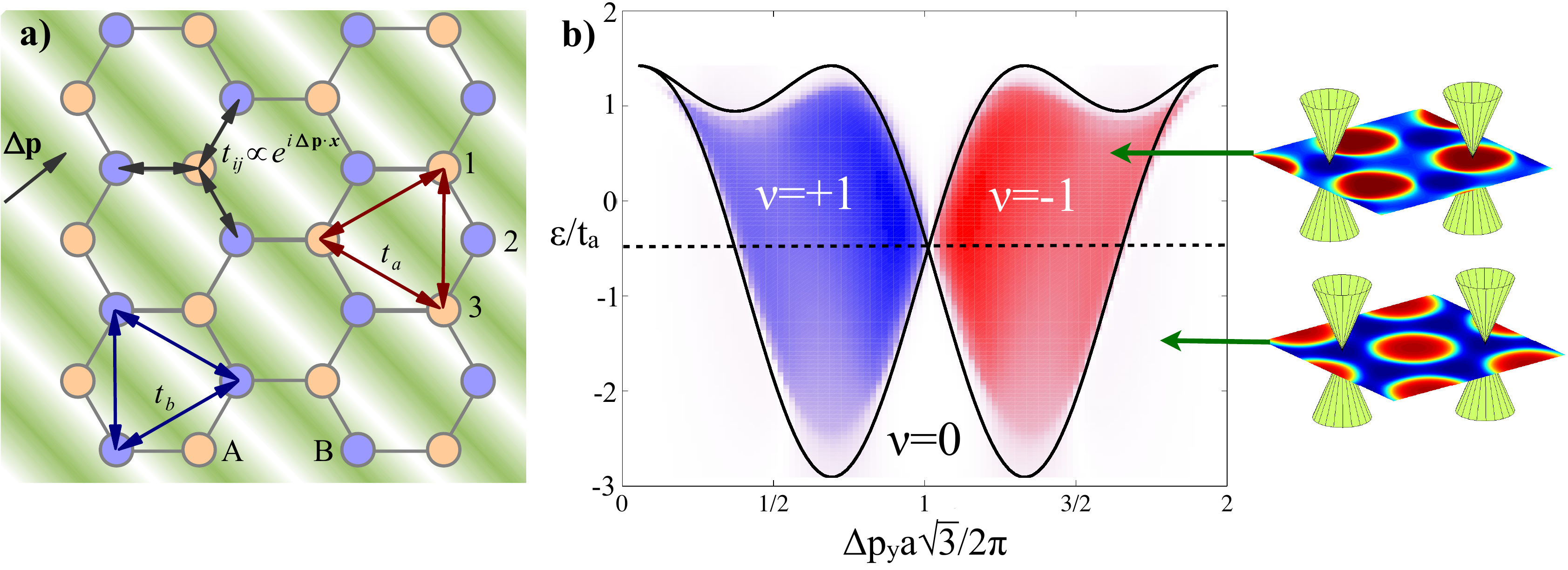}
  \caption{Haldane-type model. (a) Two triangular optical lattices (A and B) are Raman coupled by a laser that makes an atom switch sublattice ($A \leftrightarrow B$). This allows for both next-nearest neighbor hoppings, $t_{a,b},$ and a complex nearest neighbor hopping, $t_{jk}$, whose phase depends on the momentum imparted by the Raman laser, $\Delta{\bf p}.$ (b) Phase diagram of the zero-energy ground states, as a function of the energy imbalance between lattices, $\varepsilon/t_a,$ and the momentum imparted by the laser, $\Delta{\bf p}=(0,\Delta p_y).$ We plot the exact phase boundary in the thermodynamic limit (black solid line), together with a color graded simulation of the Chern number for a finite lattice with $2\times17\times17$ sites and $20\times20$ pixels. The diagrams on the right hand side show how the Dirac points are displaced on the distribution of Bloch vectors $S_z(\mathbf{k})$ induced by the Hamiltonian (blue negative, red positive).}
  \label{figlattice}
\end{figure*}

Different phases of matter can be distinguished by their symmetries. This information is usually captured by locally measurable order parameters that summarize the essential properties of the phase. Topological insulators are materials with symmetries that depend on the topology of the energy eigenstates of the system~\cite{kane05a}. These materials are of interest because they give rise to robust spin transport effects with potential applications ranging from sensitive detectors to quantum computation~\cite{Nayak,hasan10}. However, direct observation and measurement of topological order has been up to now impossible due to its non-local character. Instead, experiments have relied so far on indirect manifestations of this order, such as edge states and the quantization of conductivity.

Ultracold atoms facilitate the implementation of artificial gauge fields~\cite{dalibard10}. Here we distinguish proposals that generate continuous fields~\cite{juzeliunas04}, such as the recent experiment by Lin et al~\cite{lin09}, from those that rely on optical lattices and engineering of hopping~\cite{jaksch03}. We will concentrate on the latter, introducing a method based on standard time-of-flight (TOF) measurements that can identify a topological character in the quantum state of the system. Our starting point is a possible implementation of Haldane's model using fermionic atoms in two internal states. The topological nature of its ground state is witnessed by the Chern number. This number counts the times the ground state, written as a spinor, wraps around the sphere, as a function of momentum. We demonstrate that TOF measurements reconstruct the Chern number in a way which is robust against the presence of external perturbations or state preparation. Our method can be adapted to other quantum simulations of topological order in optical lattices~\cite{Bermudez09,Kubasiak10,Bermudez10,goldman10,Pereg11,Beri11,Bermudez11,DasSarma10,DasSarma11,Sun11}, as many already use internal degrees of freedom of the atoms to encode the order.

One common mechanism for the appearance of topological order is based on the topology of the eigenstate manifolds. Consider a real-space lattice whose unit cell has $d$ quantum degrees of freedom ---position of the particle, spin, etc---. Its energy band description, $E_m({\bf k}),$ has $d$ eigenstates, $\psi_{m=1\ldots d}({\bf k})$, per value of momentum ${\bf k}=(k_x,k_y)$ in the Brillouin zone, $\mathcal{B}$. Different configurations of the vector fields $\mathbf{v}_m = i \langle \psi_m({\bf k}) | \NABLA_{k} \psi_m({\bf k})\rangle$, defined as the gradient in momentum space of the wavefunctions $\psi_m,$ are characterized by topological invariants such as the Chern number of each band~\cite{hasan10},
\begin{equation}
\nu_m = \frac{1}{2\pi}\int_{\mathcal{B}} \NABLA_k \times \mathbf{v}_m\,{d}^2k.
\label{Chern}
\end{equation}
In the case of the quantum Hall effect, the energy bands are separated from each other and the material becomes an insulator for appropriate Fermi energies, $E_F.$ In a real setup, with finite boundaries, the sample can have a quantized non-zero conductivity given by the topological invariant $\sigma_{xy}^m = e^2/h \sum_{E_m < E_F} \nu_m ,$ which is a signature itself of topological order~\cite{VolovikBook}. The transport is then supported by ``edge'' states that are localized on the boundary of the material, with interesting properties, such as spin polarization or chirality, insulation from noise and resilience to perturbations~\cite{hasan10}. 

There are two essential routes towards topological order in momentum space, depending on how we realize the $d$ quantum degrees of freedom mentioned above. One is to start from charge carriers with intrinsic angular momentum and a spin-orbit coupling, as in the Kane model~\cite{kane05a} or in semiconductor structures~\cite{bernevig06}. A different approach is exemplified by the Haldane model~\cite{haldane88}, built on a honeycomb lattice where the unit cell has two sites and a spatially modulated magnetic field breaks the parity. We shall propose a generalization of Haldane's model that relies on optical lattice technology and two atomic hyperfine levels (pseudospin) to distinguish between the sites of the unit cell. This enables us to directly extract the Chern number from the spin textures in TOF images~\cite{koehl05} and demonstrate the topological order.

Let us consider a honeycomb lattice constructed out of two triangular sublattices, A and B [Fig.~\ref{figlattice}a]. Each of the sublattices hosts fermionic alkali atoms in a different internal state, $\ket{a}$ and $\ket{b}$. The model is parameterized by four couplings: the hopping amplitudes inside the same species lattice, $t_a$ and $t_b$, the energy difference between A and B sublattices, $\varepsilon$, and the coupling between sublattices, $t_{jk}$, which can be induced by a Raman laser and controlled at will~\cite{Eckholt09}. If the lattice is deep enough and the tunneling amplitudes and interaction energies remain small compared to the interband separation, we may use single band tight-binding model~\footnote{The neighbor relations are expressed on the honeycomb lattice and the labels $i$ and $j$ run over the unit cell indices.}
\begin{eqnarray}
  \label{eq:tight-binding}
  H =&&\!\!\!\!\! \sum_{\langle a_i,b_j\rangle} (t_{ij} b_i^\dagger a_j + t_{ij}^\star a^\dagger_i b_j)
  + \sum_{j} \varepsilon (a_j^\dagger a_j - b_j^\dagger b_j)\nonumber\\
  && \!\!\!\!\!+ \sum_{\langle\langle a_i,a_j\rangle\rangle} t_a a_i^\dagger a_j +
  \sum_{\langle\langle b_i,b_j\rangle\rangle} t_b b_i^\dagger b_j.
\end{eqnarray}
In the presence only of the hopping $t_{ij}$ the energy spectrum consists of two energy bands that meet at two ``Dirac points''. At half filling the low energy physics of the system is dominated by the linear dispersion around these points, the ``Dirac cones''~\cite{Wallace}. Due to the presence of $t_{a},t_b$ and $\varepsilon,$ the effective Dirac fermions acquire a mass that depends weakly on momentum. The position of the Dirac points on the mass landscape determines whether the model is topologically ordered or not. In our cold atoms simulation this is controlled using the Raman lasers to attach a phase to the hopping~\cite{jaksch03}
\begin{equation}
  t_{jk} \sim t \exp(i\phi_{jk}),\quad \phi_{jk}= \Delta\mathbf{p} \cdot({\bf x}_j + {\bf x}_k) / 2.
\end{equation}
This phase displaces the energy bands created by the Raman hopping, $t,$ relative to mass landscape generated by the other contributions, $t_{a,b}$ and $\varepsilon$, as shown in Fig.~\ref{figlattice}b. When the Dirac points have opposite signs of the mass, the Chern number (\ref{Chern}) automatically becomes non-zero. Intuitively, while the total flux over each hexagonal plaquette is zero, the bipartite nature of the lattice allows the phases $\phi_{jk}$ to have a non-trivial effect: along the path $1\rightarrow 2\rightarrow 3\rightarrow1,$ depicted in Fig.~\ref{figlattice}a the local effective magnetic flux, $\phi_{12}+\phi_{23},$ is also different from zero.

The momentum space Hamiltonian associated to (\ref{eq:tight-binding}) has the structure given in the introduction~\cite{hasan10}
\begin{equation}
  \label{eq:3}
  \mathcal{H}({\bf k}) =\chi(\mathbf{k})\openone -E(\mathbf{k})\, {\bf S}({\bf k}) \cdot \SIGMA,
\end{equation}
with two energy bands, $\pm{E(\mathbf{k})},$ the Pauli matrices $\SIGMA=(\sigma^x,\sigma^y,\sigma^z)$  and a normalized pseudospin $\mathbf{S}(\mathbf{k})$ labeling the state of the atoms in the $\{\ket{a}, \ket{b}\}$ space. In our model ${\bf S}({\bf k}) \propto (t\, \mathrm{Re}f({\bf k}-{\bf \Delta  p}), t\, \mathrm{Im}f({\bf k}-{\bf \Delta p}), \varepsilon + (t_a-t_b) g({\bf k}))$, with the complex functions $f({\bf k}) = \sum_{n=0,1,2} e^{-i{\bf k}\cdot {\bf v}_n a}$ and $g({\bf k}) = \sum_{n=3,4,5} \cos({{\bf k}\cdot {\bf v}_n} a)$, a set of displacements ${\bf v}_i \in \frac{1}{2}\times \{(-2,0), (1,\sqrt{3}),  (1,-\sqrt{3}), (0,2\sqrt{3}),  (3,\sqrt{3}),  (3,-\sqrt{3})\}$ and the honeycomb lattice spacing $a$. The energy shift $\chi=(t_a+t_b)g$ does not affect the topological phase.

\begin{figure}[t]
  \centering
  \includegraphics[width=\linewidth]{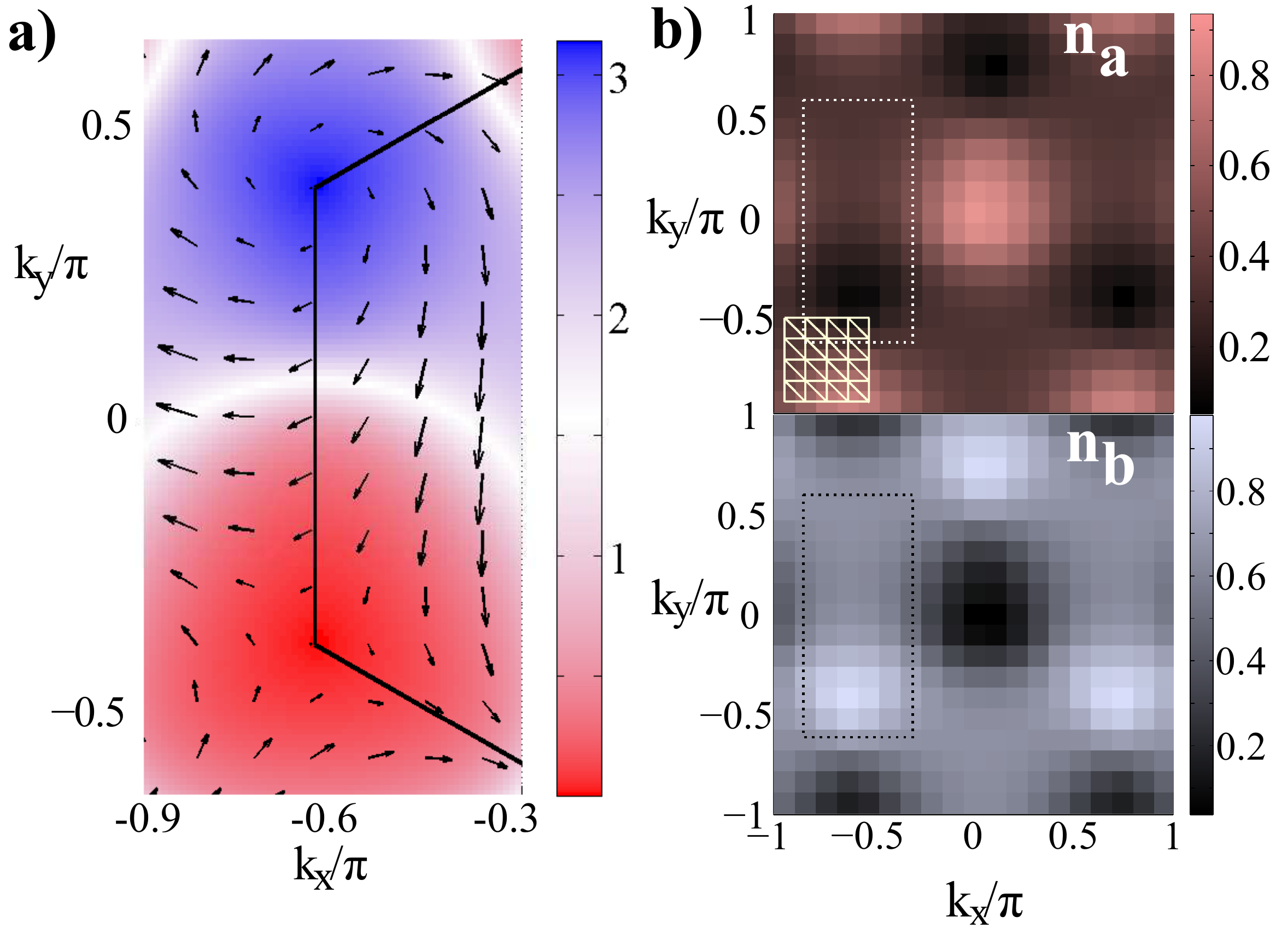}
  \caption{(Color online) (a) Spin texture of the Haldane model, interpreted as a mapping from momentum space, $(k_x,k_y),$ onto the Bloch sphere, ${\bf S}=\langle\SIGMA\rangle\propto (\cos\phi\sin\theta,\sin\phi\sin\theta,\cos\theta).$ The colors and arrows show the polar angle $\theta({\bf k})$, and the azimutal component of the spin, $(S_x,S_y)$, for a topological phase $\nu=+1$. (b) Interference pictures for this phase which result in momentum density distributions for the $a$ (up) and $b$ (down) particles. A square lattice of $20\times 20$ pixels partitioned into triangles (yellow) is used to compute the estimate $\nu_D=0.9$ [Eq.~(\ref{nuD})]. The enclosed area corresponds to (a).}
  \label{figangles}
\end{figure}

The topological properties of the model can be obtained from the field ${\bf S}(\mathbf{k}).$  In particular, the lowest energy band has a total Chern number
\begin{equation}
  \nu = \frac{1}{4\pi}\int_{\mathcal{B}} {\bf S} \cdot \left(\partial_{k_x} {\bf S}\times
    \partial_{k_y}{\bf S}\right) d^2{ k}.\label{eq:chern}
\end{equation}
Fig.~\ref{figlattice}b summarizes the three different phases that can be accessed by means of the effective magnetic flux, $\Delta\mathbf{p}=(0,\Delta p_y),$ and the imbalance between lattices, $\varepsilon/t_a.$ First of all we find a trivial region, $\nu=0,$ which is topologically equivalent to graphene with a mass term. When we interpret the associated spin texture as a map onto the Bloch sphere, both cones have the same effective Dirac mass and point to the same pole, $S_z > 0.$ Since they cover the same polar cap in opposite senses, $\nu = \pm(\frac{1}{2} - \frac{1}{2})=0.$ Moving across the solid black line in Fig.~\ref{figlattice}b, the lattice undergoes a quantum phase transition into a topologically non-trivial phase. Now the cones on inequivalent Dirac points are positioned at opposite poles of the Bloch sphere, forming a skyrmion~\cite{VolovikBook}~[Fig.~\ref{figangles}a] that covers the whole sphere and thus $\nu= \pm(\frac{1}{2}+\frac{1}{2}) = \pm 1$.

The setup in Fig.~\ref{figlattice}a may be experimentally realized along the lines of Ref.~\cite{alba11}, combining spin-dependent potentials~\cite{mandel03,jaksch98}, with recent techniques for creating dipole traps using microscope objectives~\cite{bakr09}. We suggest projecting two triangular lattice patterns on a two-dimensional sheet of light that traps the fermionic atoms. An electro-optic phase modulator controls the relative displacement of the lattices~\cite{soderberg09} and the appropriate weights of left and right circularly polarized light~\cite{mandel03,jaksch98}. The result is two hyperfine ground states of the same fermionic species confined on the two triangular sublattices of the honeycomb pattern. Thanks to this configuration, the distribution  ${\bf S}({\bf k})\propto\langle{\SIGMA}\rangle$ can be experimentally determined from the TOF images that appear when the atoms are released from the optical trap. A typical experiment would begin with a Mott state in which only the A sublattice is filled, and adiabatically progress to larger values of $t_a,\,t$ and $\varepsilon$. Once the approximate ground state is prepared, switching off the trap in adequate timescales~\cite{koehl05} projects the atom cloud into the momentum density distributions, $n_{a,b}(\mathbf{k}),$ giving direct access to one of the pseudospin components $S_z(\mathbf{k})=\frac{1}{2}[n_a(\mathbf{k}) - n_b(\mathbf{k})]/[n_a(\mathbf{k}) + n_b(\mathbf{k})]$. A fast Raman pulse during TOF allows us to rotate the atomic states and map $S_x$ and $S_y$ to $S_z$, reconstructing the whole vector field. Actual experiments ``pixelize'' the time of flight images, counting the number of atoms on each ``square'' of the effective Brillouin zone and estimating the averages of $S_x,\,S_y$ or $S_z$. Either through repetitions or through self-averaging in an experiment with multiple copies of the lattice, we will obtain a set of normalized vectors $\{\mathbf{S}_{m}\}_{m=1}^{L\times L},$ evenly sampled over momentum space. As shown in Fig.~\ref{figangles}b, we suggest identifying the pixels with the nodes of a triangular lattice, $T=\{\mathbf{S}_{j_T},\mathbf{S}_{k_T},\mathbf{S}_{l_T}\}$, approximating the integral $\nu$ by its discretization
\begin{equation}
\label{nuD}
\nu_D := \frac{1}{8\pi} \sum_T {\bf S}_{j_T}\cdot{\bf S}_{k_T}\times{\bf S}_{l_T}
=\nu + {\cal O}(4\pi^2/L^2).
\end{equation}
The value $\nu_D$ has the properties of a topological quantity ---stability and robustness against local perturbations---, and is also stable with respect to the discretization~\cite{fukui05}.

\begin{figure*}[t]
  \centering
  \includegraphics[width=\linewidth]{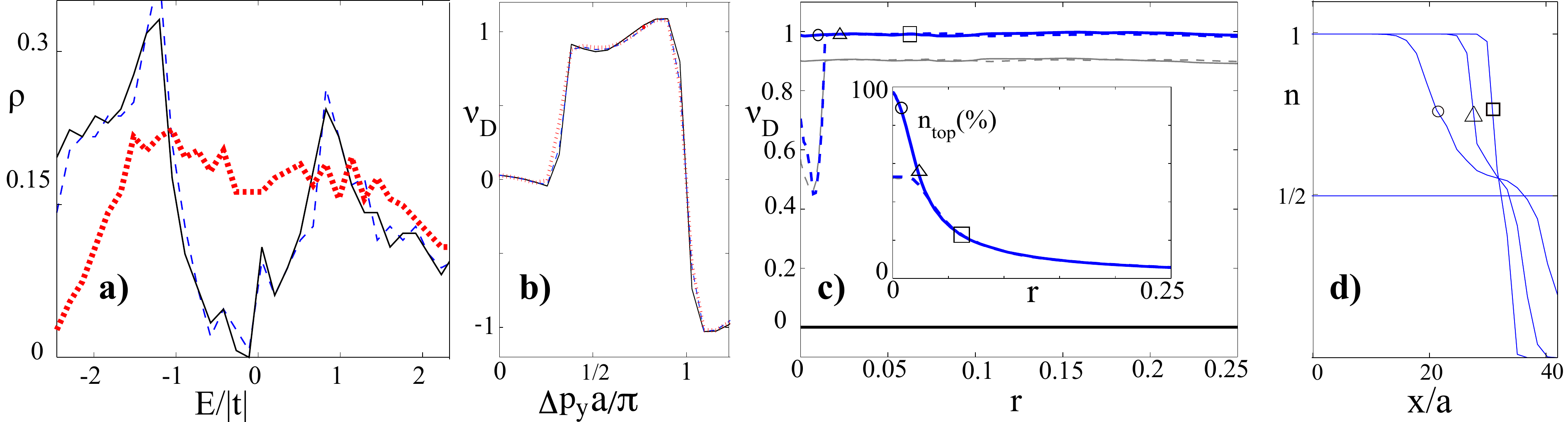}
 \caption{(Color online). (a) Density of states and (b) Chern number simulation at $\varepsilon/t_a=-0.5$ for a lattice with $r=0$ (black), $10^{-3}$ (dashed) and $0.02$ (dotted). (c) Chern number for $\varepsilon/t_a=-0.5$ and $p_y=3\pi/4a$ (blue, above) or $2\pi/a$ (black, below) vs. confining potential strength, $r$, starting with $1/2$ (solid) or $1/4$ particles per site (dashed) on a lattice with $50\times 50$ sites. The values obtained using $50\times50$ pixels are compared with those from a $20\times20$ matrix (gray). (d) Density plots for the points marked in (c), showing the wedding cake structure where regions with $n=1$ do not contribute to the Chern number.}
  \label{FigSim}
\end{figure*}

We have compared the thermodynamic limit distribution $\mathbf{S}(\mathbf{k})$ with realistic finite-size lattices with imperfections. For this we have exactly diagonalized Eq.~(\ref{eq:tight-binding}) on a finite lattice with up to $20000$ sites, including the additional harmonic confinement term, $\frac{1}{2}m \omega^2 x_i^2,$ which is typical from cold atom experiments. Our plots report simulations with $t/\hbar=1$ kHz, $t_a/t=0.5$ and $t_b=0$, using $r=m \omega^2 a^2/2t$ to parameterize the influence of the harmonic confinement. Realistic values for a lattice with $a\sim 400$nm range from $r=10^{-3}$ ($^6$Li in a trap with $\omega/2 \pi=60$ Hz) to $r=0.02$ ($^{40}$K in a trap with $\omega/2 \pi=100$ Hz), but we probed up to $r=0.25$. The results are very insensitive to the number of atoms, as already for $17\times17$ sites the interference pattern provides the right phase diagram [Fig.~\ref{figlattice}b]. The Chern number is also very robust with respect to the discretization: a $20\times20$ pixelization deviates from the theoretical value of $\nu$ only $10\%$~[Fig.~\ref{FigSim}b], in line with the error ${\cal O}(4\pi^2/L^2),$ expected from a discretization with a smooth integrand. Moreover, $\nu_D$ still captures the discontinuity across the topological phase transition [Fig.~\ref{figlattice}b]. Contrary to the global density of states and the eigenenergies, the approximate Chern number $\nu_D$ is also robust against inhomogeneities. Already for a confining trap with $r = 0.02$ the Dirac cones are no longer evident~[Fig.~\ref{FigSim}a], but the Chern number is still close to $\pm 1$ with a good signal-to-noise ratio ~[Fig.~\ref{FigSim}c]. This is due to the wedding cake structure introduced by the harmonic trap~\cite{Zhu07}: for sufficiently strong traps there is always one ring or disc hosting $n_{\mathrm{top}}$ atoms in a topological phase [Fig.~\ref{FigSim}c-d]. Only these atoms contribute to the total Chern number, much like only superfluid atoms add to the interference peaks in experiments with bosons in optical lattices~\cite{foelling06}. Note also how, as shown in Fig.~\ref{FigSim}d, for low densities there are not enough atoms to form a topological phase and the Chern number deviates from $\pm 1$. However, raising the trap brings the chemical potential up to a level in which the first disc with particles in a topological phase is created, and $\nu_D$ converges to $\pm 1$. Finally, we expect also a good behaviour in finite-temperature simulations because the effect of temperature only changes the length of vector $\langle\SIGMA\rangle$ (i.e. the signal), but not its orientation $\mathbf{S}.$

Summing up, we have presented a robust and very general method to detect topological order in momentum space using ultracold atoms in various internal states and TOF images. As a very relevant application we have introduced an experimental proposal to generalize the Haldane model~\cite{haldane88}.  We found that the topological phases and our method are both robust under (i) use of small finite lattices (ii) coarse grain measurements of the spin texture, (iii) inhomogeneous potentials superimposed on top of the lattice, and (iv) errors in the exact values of the chemical potential, number of atoms or finite temperature. We believe our proposal is thus advantageous with respect to other indirect detection schemes ---edge transport, eigenstate preparation~\cite{goldman10} or local estimates of the density of states~\cite{Zhu07}---, which may be more sensitive to temperature and imperfections. Compared also to the numerical protocol in Ref.~\cite{ringel11}, our method only requires a single set of measurements instead of reconstructing spatially dependent correlators in position space, which are not easily accessible in optical lattices. Finally, the implementation of our ideas would represent the first direct visualization of non-local topological order.

This work has been funded by Spanish MICINN Project FIS2009-10061, FPU grant No.AP 2009-1761, CAM research consortium QUITEMAD S2009-ESP-1594, a Marie Curie Intra European Fellowship, JAE-INT-1072 CSIC scholarship and the Royal Society.

%\bibliographystyle{apsrev}
%\bibliography{insulator}

\end{document}